%% file: pck-paper.tex
\documentclass[prb,preprint]{revtex4-1} 

\usepackage{multirow} 
\usepackage{amsmath}  
\usepackage{amsfonts} 
\usepackage{graphicx} 
\usepackage[breaklinks]{hyperref} 
\newcommand{\etal}{\textit{et al. }}

\begin{document}


\title{A collaboration to support novice instructors in research-based astronomy teaching}

\author{Sissi L. Li}
\email{sili@fullerton.edu}
\affiliation{Catalyst Center, California State University Fullerton, Fullerton, CA 92831 USA}

\author{Gabriela Serna} 
\affiliation{Department of Physics, California State University Fullerton}
\affiliation{Gravitational Wave Physics and Astronomy Center, California State University Fullerton, Fullerton, CA 92831 USA}

\author{Heather Chilton}
\affiliation{Department of Physics, California State University Fullerton}

\author{Michael E. Loverude}
\affiliation{Department of Physics, California State University Fullerton}
\affiliation{Catalyst Center, California State University Fullerton, Fullerton, CA 92831 USA}

\author{Jocelyn S. Read} 
\affiliation{Department of Physics, California State University Fullerton}
\affiliation{Gravitational Wave Physics and Astronomy Center, California State University Fullerton, Fullerton, CA 92831 USA}

\author{Joshua R. Smith}
\affiliation{Department of Physics, California State University Fullerton}
\affiliation{Gravitational Wave Physics and Astronomy Center, California State University Fullerton, Fullerton, CA 92831 USA}
\email{email@email.edu} 


\date{\today}

\begin{abstract}
\input{abstract.tex}
\end{abstract}

\maketitle 

\section{Introduction} 
\label{sec:intro} 
\input{intro.tex}

\section{Context}
\label{sec:context}
\input{context.tex}

\section{Revised Course}
\label{sec:program}
\input{program.tex}

\section{Process of Faculty Development}
\label{sec:documenting}
\input{documenting.tex}


\section{Results of Faculty Development}
\label{sec:results}
\input{results.tex}

\section{Assessment of Student Learning}
\label{sec:assessment}
\input{assessment.tex}

\section{Conclusions and Ongoing Work}
\label{sec:conclusions}
\input{conclusions.tex}


\begin{acknowledgments}
\label{ack}
\input{ack.tex}
\end{acknowledgments}

\bibliographystyle{unsrt}
\bibliography{references}

\end{document}

%% file: abstract.tex
Introductory astronomy is one of the most widely-taken university science courses and a common means of satisfying general education requirements.  This course is frequently taught in large sections by a variety of instructors, some with little formal training in astronomy.  Thus, instructors often struggle to help students connect with what is typically their last formal science experience. At California State University Fullerton, we have engaged in reform efforts in  the teaching of introductory astronomy. We report on the model of reform implementation with several key features that we believe were instrumental in our success. First, our astronomy instructors learned to use research-based curricular materials by attending workshops before and after first implementation. Second, the astronomy instructors collaborated with student Peer Instructors as a team to document and reflect on the implementation progress. Third, the instructional team of faculty and Peer Instructors worked with science education researchers to collect information on common student conceptual difficulties, and develop strategies to address these difficulties using resources in the research and teaching literature. Our findings suggest that this is a fruitful and sustainable model for reform implementation, faculty professional development, and research on the teaching and learning of astronomy.




%% file: intro.tex


In recent years there has been a renaissance and revolution in the teaching of introductory astronomy to help students learn in more meaningful ways. This movement is due in part to the influence of discipline-based education research~\cite{singer2012} (DBER), primarily astronomy education research~\cite{prather2009, rudolph2010} (AER) but also its close cousin physics education research~\cite{meltzer2012} (PER). Through the work of researchers and curriculum developers in these fields, a variety of interactive methods and curricular materials have become available to instructors. 

At the same time, there is considerable evidence that there are barriers, real or perceived, that inhibit faculty from adopting and using DBER-based curriculum materials~\cite{henderson2007}. Many instructors who try research based instructional strategies discontinue their use due to frustration or modify them dramatically so that the strategies are no longer faithful to the original goals~\cite{henderson2009}. In particular, researchers found that while 72\%  of university faculty sampled have tried to implement RBIS at least once, about a third of these faculty stop after a single attempt due to lack of success and support~\cite{henderson2012}. Henderson \etal assert that ``successful strategies provide support during implementation in the form of performance evaluation and feedback''~\cite{henderson2011}.
 
So while resources for interactive learning, and assessment, for introductory astronomy courses exist, we assert that instructors require additional support to successfully and sustainably implement course reform. We will present a successful implementation of research-based reform and a systematic study of the elements that supported instructor persistence, enthusiasm, and development. Important elements of support include workshops to learn about the DBER-based instructional strategies, a reform implementation team of instructors and researchers, and ongoing reflection and research during the implementation.

This paper focuses on the efforts to reform the astronomy course at one institution and implement research-based instructional materials. Our goal is to present a model of faculty development for junior university faculty that effectively incorporate research-based instructional strategies, team-based teaching, and action research. We will first describe the existing course and the motivation for change. We will then describe the professional development of the faculty and peer instructors to describe the reflective process of course reform. In particular, we will examine instructional choices related to instructor awareness and understanding of student difficulties and ideas about how to learn astronomy. We have collected a variety of qualitative and quantitative data to document the impact on student learning including conceptual inventories, classroom observations and instructors' reflective journals. We will demonstrate that improvements in student conceptual understanding during the implementation course reforms were excellent compared to similarly interactive astronomy courses.  

\subsection{Course Reform and Faculty Development}

Many interactive methods and curricular materials have been developed and evaluated by researchers in the discipline-based education research fields. Because these teaching strategies and curricula are available to instructors, course reforms have been implemented with varying success in science classrooms from K-12 to higher education.  Examples of widely adopted research-based instructional strategies include Peer Instruction~\cite{mazur1997}, tutorials~\cite{mcdermott2002}, and Lecture-Tutorials~\cite{prather2004}.  With the formal study of student learning in astronomy came documentation of student conceptual difficulties and the development of standardized assessment instruments covering typical course content~\cite{bardar2007,bailey2007}. Among the barriers are departmental norms, student resistance, and lack of instructor time. In particular, Henderson and Dancy noted that ``instructors are sometimes too busy with large teaching loads and/or research responsibilities to have the time to learn about and integrate new techniques~\cite{henderson2009}." 
 
  
 \subsection{Pedagogical Content Knowledge}
In order to characterize some of the essential learning of the novice instructors in this study, we have turned to literature on teacher preparation. In addition to the challenges in adopting curricular reform in a productive manner, there is evidence to believe that new instructors often lack Pedagogical Content Knowledge (PCK): the knowledge that experienced teachers bring to bear on the teaching of their discipline. Shulman\cite{shulman1986} defined PCK as a category of knowledge that is distinct from knowledge of the facts and theories of a discipline (content knowledge) as well as from curricular materials and the procedural use of these ``tools" for teaching (curriculum knowledge). Instead, PCK is a discipline-specific set of knowledge that teachers develop of how to approach their discipline from the students' point of view, anticipate student struggles, and understand student thinking~\cite{ball2008}. Much of the research on understanding and measuring PCK has been developed in math education research community \cite{Hill2008} and has focused on the K12 levels \cite{Loughran2004}. It is challenging to study systematically PCK and its development in part due to the difficulty in clearly defining what constitutes PCK and what markers indicate a change in PCK. Further, because PCK is specific to a course and population, we must be cautious in translating the results from another discipline to applications to astronomy teaching. For our study, we examined markers in teaching that have been associated with PCK development rather than directly assessing the PCK development of instructors. 

For a new university faculty member whose graduate training is almost exclusively focused on research, the development of PCK can be a daunting task.  New instructors assigned to introductory astronomy courses typically have a strong background in physics and/or astronomy, however they often have little PCK. Further, their primary focus is on ensuring that they have mastered the astronomy content knowledge, understanding how to use the curriculum, and managing the classroom; as a result, there is little room left to reflect on and actively develop PCK. However, a lack of PCK can make it very difficult for instructors to recognize how curricular goals are reflected in practice and predict how the students will respond to curricular elements. Because implementation of course reforms can be highly sensitive to a local context, instructors with well developed PCK are better able to troubleshoot and tailor the curriculum to the class and student needs.  While there are resources available to instructors (such as workshops, modules), many new instructors are not aware of the need to develop their PCK. 

At our institution, our ongoing reform efforts in astronomy courses have strong departmental support and close collaboration with science education researchers. As a result, we have a context well suited for developing PCK; further, we can examine the ways in which instructors may develop PCK. We believe this research will provide insights about how instructors develop PCK, specifically how they learn to address specific student difficulties in learning astronomy. In studying the development of PCK, we have developed a model of small-scale faculty development that may be useful for other departments seeking to adopt reformed instructional strategies and to help instructors to develop pedagogical skills and PCK.  

 \subsection{Epistemological Expectations}

The initial focus of our work was on content knowledge and content-oriented PCK.  However, we recognized that many of the issues in day-to-day instruction and in the peer instructor's field notes focused not on content understanding but rather on students' frustrations about class procedures.  When we considered these issues, they appeared to reflect conflicts with what some researchers have described as students' epistemological expectations~\cite{Redish1998, hammer1994}.

Redish \etal describe epistemological expectations as follows:  ``Each student brings to the …class a set of attitudes, beliefs, and assumptions about what sorts of things they will learn, what skills will be required, and what they will be expected to do.”  Following earlier work by Songer and Linn~\cite{songer1991} and Hammer~\cite{hammer1994} they characterize student beliefs about learning in terms of six dimensions:  independence, coherence, concepts, reality link, math link, and effort, and identify favorable and unfavorable perspectives for each of these dimensions.  For example, in the 'concepts' dimension, a student with a favorable perspective would focus on developing deep conceptual understanding, and a student with an unfavorable perspective would concentrate on memorization. 

While these non-content issues about learning frustrations can be attributed to mismatches in student and instructor expectations about how to learn, the issues may also arise due to other reasons such as affective issues and cultural expectations. For example, the instructional motivation for having students discuss their ideas and solve problems in groups is that the development of understanding is shared and meaning is constructed socially.  This can be an epistemological mismatch if, for example, students believe they are supposed to learn by figuring things out individually or that they should passively accept facts from an authoritative instructor. Resistance towards group work can also be an affective issue in that the student finds the other students disrespectful and competitive. 

While we do not have sufficient detail or quantity of data to determine the cause of non-content frustrations in our course reform, we use the construct of epistemological expectation as an indicator of situations where non-content pedagogical knowledge for teaching in an interactive social learning classroom might be needed and developed. Awareness and understanding of this mismatch of expectations can facilitate more successful reform efforts, help students develop favorable epistemology and support instructor development.


%% file: context.tex
\subsection{The local setting}

California State University Fullerton (CSUF) is a large comprehensive university in Orange County in Southern California.  CSUF serves a very diverse student population of over 38,000 students and is one of the largest universities in the state.  At CSUF, the introductory astronomy course for non-science majors has traditionally been taught as PHYS120, Introduction to Astronomy.  This course satisfies the general education requirement for scientific inquiry and quantitative reasoning in physical science, which is required for all CSUF bachelor's degrees.  At the beginning of the project described in this paper, PHYS120 was a stand alone course with no laboratory, usually taught in lecture sections of 50 to 100 students.  The course typically used a standard introductory textbook~\cite{seeds2012} and was largely traditional.  It was taught by a mix of faculty, including one tenured faculty member who is a research astronomer and several part-time faculty with various non-astronomy backgrounds, typically an MS or PhD in physics.

Since 2011, driven in part by a collaborative effort between new faculty members specializing in gravitational-wave astronomy and the Catalyst Center for Research in Science and Math Education, there has been an increased interest in teaching astronomy using reformed instructional approaches based on AER. During this period, the department's emphasis on astronomy grew as well, with initiatives including the re-introduction of a laboratory component to introductory astronomy, the re-naming of the introductory course from PHYS120 to ASTR101, and the planned introduction of elective courses for physics majors and potential astronomy minors. 

\subsection{The research/teaching team}

This project was carried out by a collaborative team of education researchers, traditional physics and astronomy faculty, and peer instructors. The instructors for this course are both junior faculty members who taught the course for the first time as part of this project. Smith, a researcher in experimental gravitational-wave astronomy and optics, had taught two sections of calculus based introduction to mechanics at CSUF prior to teaching PHYS120. For Read, a neutron-star and gravitational-wave astrophysicist, PHYS120 was the first course she taught as a faculty member. The peer instructors Serna and Chilton are both physics majors who had taken an introduction to astronomy course before (although not one taught by Smith or Read). They received independent study credit for their participation. Researchers Li and Loverude provided support by establishing a reflective journaling process, advising in the collection of research data and classroom observations, and guiding the faculty and peer instructors to reflective upon their PCK. Serna and Chilton both took detailed field notes of the lecture classes in which they were peer instructors. Serna took the lead role in the collection and analysis of the quantitative data described in Section VII.

The new faculty who would teach the revised astronomy courses were enthusiastic and had some familiarity with curricular innovations  based on physics and astronomy education research.  At the same time, they had relatively little prior teaching experience, and in particular little prior experience with either introductory astronomy or interactive engagement instructional strategies.  We characterize these new faculty as being very strong in astronomy content knowledge but lacking in pedagogical content knowledge and in experience establishing classroom social norms that support interactive engagement. 

%% file: program.tex
As previously noted, the modifications to the Introduction to Astronomy were motivated by the fact that it is one of the most widely taken science courses in the US and, for most students, their last formal experience with any science. As such, the course provides an opportunity to raise student interest in science, to encourage students to understand the world as a place in which events happen based on predictable causes, and to help students understand the science behind situations from their daily experiences, that they might otherwise have seen as random.

The revised course was set up with this in mind. The goals of the course are that students should learn to, 
\begin{itemize}
\item understand the nature of science through the eyes of astronomy,
\item understand the big ideas in astronomy, and
\item develop or deepen a lifelong interest in astronomy. 
\end{itemize}

Education research suggests that large lecture courses fail to engage most students and that learning science concepts is more challenging than most instructors imagine. Therefore, to better achieve these goals we have adopted a reformed instructional format. Although the course still meets in sections of between 30 to 100 students, instructors spend about a third of the class time using active learning strategies.   

The course topics are largely unchanged. They include motion of the night sky and the celestial sphere, moon phases, the development of modern astronomy with Newton's and Kepler's laws, light and spectroscopy, and the life and death of stars. Because of their area of research specialization, the instructors also devote a small amount of time to gravitational-wave astronomy. 
 
The tools and methods used to increase classroom interactivity were adopted largely from the NASA Center for Astronomy Education (CAE). Smith and Read were introduced to these materials largely through CAE workshops~\cite{CAEhilo, CAEanchorage} and interactions with other practitioners, as described in the next section. 

Each 75-minute class period consisted of a number of instructional activities. Several mini-lectures, 10-20 minutes in length, served to introduce key concepts and often included simulations and/or lecture demonstrations. These were punctuated by conceptual think-pair-share (TPS) questions~\cite{green2003,slater2003}. Students responded with A-B-C-D cards and the questions were debriefed by asking the students to answer together aloud (reinforcing that the correct answer is well known by the students), or to discuss with each other about the correct answer. Roughly once per class, students engaged in more extended group activities such as Lecture-Tutorials~\cite{prather2012lecture} and ranking tasks. Weekly work outside of class included roughly ten pages of reading prior to class from Bennett's The Essential Cosmic Perspective~\cite{bennett2012}, and one homework set with roughly ten questions through the online system MasteringAstronomy.

Prather \emph{et al.}~\cite{prather2009} introduced a metric they called ``Interactivity Assessment Score (IAS)," in order to quantify what portion of class time in any given course involved student interaction. As part of this project, two instructors independently reported to Rudolph the total number of TPS questions, ranking tasks, and Lecture-Tutorials used, and he used that data to calculate IAS values ranging from 27\% to 38\%, as shown in Table~\ref{tab:lsci}. Figure~2 in Rudolph \emph{et al.}~\cite{rudolph2010} shows the correlation observed between normalized gains on the LSCI exam and the corresponding IAS values in a national study in introductory astronomy courses. Comparing to this, for the level of interactivity achieved in our courses, most of the normalized gain values are above the national average. 





To support the effective implementation of interactive teaching methods, each faculty member was joined in class by a student peer instructor. This study included four sections with full peer instructor support, two sections with peer instructor support during one of two lectures per week (because of scheduling conflicts), and one section (Read, Spring 2013) with no peer instructor, as shown in table~\ref{tab:lsci}.




%% file: documenting.tex

\subsection{Pre-teaching preparation}

Both faculty instructors took part in NASA Center for Astronomy Education (CAE) two-day workshops to enhance their pedagogical preparation and become familiar with the existing materials in astronomy education. In summer 2011, Smith participated in the CAE workshop on effective implementation of interactive teaching in astronomy~\cite{CAEhilo}. Read attended a similar workshop in January 2013. In summer 2012, Smith participated in a second CAE workshop aimed at using technology in the classroom~\cite{CAEanchorage}. The instructors were introduced to Lecture-Tutorials, ranking tasks, think-pair-share questions, and astronomy simulations and practised their implementation in mock classroom situations. 

Developing PCK was a key focus of the CAE workshops and was explicitly introduced early on.  The workshop website states~\cite{CAEworkshops}:  
\begin{quotation}
Almost all college Science, Technology, Engineering, and Mathematics (STEM) instructors, postdocs, and graduate students receive little to no formal training in how to effectively implement active-learning instructional strategies that have been proven to improve students' knowledge, skills, and beliefs beyond what is typically achieved in college STEM courses.
\end{quotation}
The goals of the workshops are described thus:  
\begin{quotation}
Our Tier I Teaching Excellence Workshops will provide you with the experiences you need to create effective and productive active-learning classroom environments. We will model best practices in implementing many different classroom-tested instructional strategies. But most importantly, you and your workshop colleagues will gain first-hand experience implementing these proven strategies yourselves. During our many microteaching events, you'll have the opportunity to role-play the parts of student and instructor.
\end{quotation}

In preparing the reformed course, Smith and Read adopted the CAE interactive materials and also drew heavily from lecture slides by Professor Duncan Brown at Syracuse University, an early adopter and CAE collaborating member. Together with the peer instructors, they worked through all of the classroom materials prior to their first courses -- and came into the classroom well prepared.  

\subsection{The role of peer instructors}

Initially, the vision was that each peer instructor would help with the management of the large lecture classes by facilitating classroom activities and interacting with students during tutorials. What emerged was that they played a much more important role than that; they could interact in a less formal fashion with the students and presented less of a barrier for students to ask questions and express dissatisfaction. Additionally, our peer instructors were dedicated multi-semester research students, and they were able to build upon their PCK in each successive semester.

During class, the peer instructor would typically sit in the audience listening to lecture recording field notes including student questions, response patterns to TPS questions and issues emerging from the tutorials. During Lecture-Tutorials, the peer instructors would join the faculty in circulating around the classroom, and both would respond to student questions, observe student written responses and listen to the small group discussions, intervening when groups appeared to be on the wrong track. Both instructors attempted to engage students first by asking questions and rarely provided direct explanations. 

With this expanded role, the peer instructors were essential in gathering information concerning PCK, particularly through taking field notes and documenting student issues and ideas about the materials. The great majority of the results presented in Section~\ref{sec:results} were gleaned from peer instructor observations. Their roles as almost-instructor and almost-student gave them insider and outsider perspectives on the course activities and student interactions. This expanded role served to help instructors critically evaluate the reform efforts and to help the peer instructors build their own PCK.  


\begin{table*}
\caption{\label{tab:data} Participation and documentation by semester. }
\small
\setlength{\tabcolsep}{0.5em}
\begin{tabular}{  l  l  l  l  }\hline
Semester & LSCI (Pre/Post) & Reflective Journals & Peer Instructor Field Notes \\
\hline
Fa2011 & Y & N & Serna\\
Fa2012 & Y & Chilton/Serna/Smith/Read  & Chilton/Serna \\
Sp2013 & Y & Chilton/Serna/Smith/Read & Chilton/Serna \\
\hline
\end{tabular}
\end{table*}

\subsection{Further pedagogical/PCK development}

The group of researchers (two instructors, two peer instructors and two science education researchers) met several times during each 15 week semester to discuss teaching developments and research data collected and analyzed. Although not used directly for research data, the instructors and peer instructors also met occasionally during the semester to align their teaching. In addition, Smith and Read share an office suite and talked regularly about PCK issues in their sections.  

Prior to the start of term, an online reflective journal was set up for the instructors and peer instructors with a general prompt to evaluate how the class went. These journals were hosted as a private website on Google so that viewing and editing were only possible with an invitation. Everyone in the group was able to view and comment on each other's journal entries. The group made use of this feature as evidenced by comments from each group member on at least one other member's entries and several instances of one person referencing another's entry. The content of the journals was not restricted to specific evaluations or topics. As a result, the entries encompassed a wide variety including the teacher's emotional state, expectations about the class activities, reaction and evaluation about the instruction, and ideas for improvement. Using these journal reflections, we were able to highlight common student struggles across courses and document different perspectives on specific classroom activities. The journals were also a valuable documentation of how the teacher's views, attitudes and ideas about teaching evolved over time. Through these learning and reflective processes, the instructors not only developed PCK for teaching astronomy, but were consciously aware of the developmental process.

The following are data we collected which are discussed in Table~\ref{tab:data}

\paragraph{LSCI:} We used the Light and Spectroscopy Concept Inventory exam (LSCI). This is a 26-question diagnostic test used to evaluate student understanding of light and spectroscopy. 

\paragraph{Reflective Journals:} Online reflective journals where the professors and the peer instructors were able to discuss teaching and relevant experiences such as:  preparation, in-class interactions, discussions with colleagues, and other teaching related experiences.

\paragraph{Peer Instructor field notes:} Notes taken by the peer instructor discussing student struggles, questions, and responses(percentage) to TPS questions.


%% file: results.tex
\subsection{Epistemological expectations}

Throughout the courses described here, many of the trouble spots that were identified by faculty and peer instructors can be characterized as stemming from discrepancies between student expectations and those of the instructors. While Shulman's PCK deals with how the instructor chooses the appropriate instructional strategy to help a student learn content, other instructional choices take into account expectations about the nature of learning. These expectations can reflect content learning, such as the role of math in learning astronomy, or other aspects of learning, such as the importance of social interaction in group work. Our data sheds light on the epistemological expectations of students and the ways in which they conflict with the expectations of instructors and the practices in reformed courses. 

We describe two examples of epistemological expectation mismatch that roughly map to the independence and math dimensions described by Redish \emph{et al.}~\cite{Redish1998}. We also describe a third example that is not reflective of Redish's work, that of addressing expectations about social norms that indirectly impact content learning. In each of these contexts, the qualitative research data from instructor reflective journals and classroom observations reveal the nature of the mismatch and the instructors' responses as a result of this awareness.

\subsubsection{Independence: Students taking responsibility for learning vs. Instructors transmitting information}

There has been an ongoing tension between a (sometimes unstated) student expectation that students will passively receive and reproduce information from the instructor and the faculty expectation that students will take responsibility for their own learning.  For example, during the third lecture of one section, a peer instructor noted:  ``More students in this section take more ``traditional notes" and appear less engaged. \dots Instead of copying the slides, [they should] write down key or problem concepts and `lightbulb' moments. Other than that, it's just too easy to get drawn into an unthinking state of copying what's written."  The students' concern with copying down the contents of lecture slides, despite the slides being readily available on the course management system, reflects the expectation that the students' job was to make sure they got the information as presented. This is clearly at odds with the peer instructor's idea that learning should include thinking about  the presented material in the larger conceptual context and not just the immediate information presented. After lecture 12, a peer instructor noted ``It was kind of funny – an exam study sheet was really wanted, a kind of outline of what to study, despite having the practice exam.  Not to mention both the Lecture-Tutorials and the clicker question slides are perfect study material, but I guess this is not obvious?" Again, the instructor expected that students to make use of the many resources for studying and take charge in organizing their study efforts. Instead, the students perceived their responsibility to be following the instructor's list of what to study so that they could reproduce the information on the exam. 

This conflict also took shape in student responses to the Lecture-Tutorials. The Lecture-Tutorials are designed to help students build their own understanding and to take responsibility for their own learning.  As a result, the correct answers are not provided.  However, there is repeated evidence that students prefer that information be given to them.  After the class period 6, a peer instructor wrote, ``One student brought up a good comment in regards to getting verification with the Lecture-Tutorials:  Sometimes a student can feel convinced, even having gone through their reasoning and be wrong.  While I still prefer not to tell students they are right or wrong and get them comfortable with their own reasoning, the student did have a point."  In this reflection, the peer instructor documented her observations and thoughts that the student expectations about problem solving that were not aligned with those of the instructor team. From this first step in developing awareness and empathy for the mismatched expectations, the peer instructor  related and discussed the issue with the instructor.

It was clear that some students did not enjoy doing tutorials, though the reasons weren't always clear.  On class period 5, when the time came for the students to begin the Lecture-Tutorial (LT), the tutorial and page numbers were announced on a slide.  The peer instructor noted, ``There were actually quite a few groans when the slide for the LT popped up."  This difficulty persisted for many students throughout the semester.  Near the end of the semester notes recorded again that there were ``Lots of groans around the classroom when the students realized they were doing a LT again."  In her end of class summary, the peer instructor wrote, ``There has been a lot of contention by the students about not being able to check their answers or going over the Lecture-Tutorials.  At least four students had written that as the major thing they disliked about the class and many more have vocalized the same throughout the Fall 2012 semester.  In the Spring 2013 semester, 18/~53 wanted LT answers and 20/~53 disliked the Lecture-Tutorials."  

To help resolve this mismatch, instructors have tried a number of responses to student concerns about not receiving answers to the Lecture-Tutorials.  After initial complaints about Lecture-Tutorials not being graded and students not knowing whether their answers were correct, one instructor collected the tutorials for grading.  There were further complaints from students about this change in classroom procedures.  The instructors modified their response further and ultimately made adjustments with little functional change to the class or instructional model but assuaged student concerns and explicitly reminded the students the point of the activity.  The class has always followed Lecture-Tutorials with a series of clicker questions that serve to review the concepts covered in the tutorial.  Now the instructors explicitly begin this sequence with language like the following:  ``I have a series of clicker questions to debrief that tutorial, and make sure we're all on the same page."  The instructors also select a key question from late in the tutorial and ask the class to answer the question orally in unison, which serves to demonstrate the new level of classroom consensus in a explicitly public and social manner. In the process of understanding the epistemological mismatch and making these changes, instructors developed better understanding of how to help students shift their ideas about how to learn astronomy. These changes add almost no time to the lecture, but the instructors have reported that there is less resistance to the tutorials.

\subsubsection{Math:  Values mathematical representations vs. Sees math as the enemy}

In the Redish \emph{et al.}~\cite{Redish1998} work this continuum is described as ranging between extremes of ``considers mathematics as a convenient way of representing physical phenomena" and ``views the physics and the math as independent with little relationship between them."  That work was conducted in the context of calculus-based physics for science and engineering students.  For the non-science majors taking Introductory Astronomy, the unfavorable end of the spectrum is probably better characterized as ``reacts negatively and/or shuts down entirely when math is required."  While the emphasis in the reformed course has been on conceptual understanding and not algebraic manipulation, the course materials and instructors do use symbolic expressions and equations as means of summarizing relationships and ask students to reason qualitatively with these relationships. It is apparent that for many students even this qualitative mathematical reasoning is not only challenging but also viewed with negativity. On the few occasions when equations appeared in the class, student responses were generally quite negative. A peer instructor wrote about the equation associated with parallax (see equation 1 in this paper) ``\dots and the moment it appeared on a slide, several students groaned and made negative comments.  How to manipulate the equation for the desired result, without any unit conversions, had to be shown to some students." The equation showed the simple inverse relationship between two variables, and the peer instructor had expected more mathematical understanding and acceptance of math as a part astronomy. In the following lectures, the peer instructor documents consistently negative students reactions to the use of equations. In lecture 10, she noted that ``Class seemed really disinterested\dots and there was some dissatisfaction with having to work with an equation."In lecture 14, ``There is definitely confusion between the gravitation force of an object as described by F=ma and the gravitational force between masses $F=GMm/r^2$.  … students seem intimidated by equations..."


\subsubsection{Expectations about classroom social norms}  

These instances reflect divergence between the expectations of students and instructors.  Such divergence is not uncommon for faculty seeking to adopt reformed instructional strategies, and there is evidence that student resistance to reformed strategies is one reason that faculty discontinue their use of such strategies ~\cite{henderson2012}.  This is particularly critical as the faculty in our project are untenured; for them the student evaluation of teaching is a significant part of their tenure and promotion process.  For this reason we feel that it is useful and important for instructors to understand and anticipate the potential conflicts arising from this mismatch and develop strategies for mitigating these conflicts.   

As noted above, the instructors who participated in the CAE workshops received some formal preparation in this regard.  The workshop devotes a significant amount of attention to anticipating student resistance to instruction and in positive and preemptive steps to ensure student buy-in.  However, our experience suggests that instructors need ongoing support.  We feel that instructors need to make their own choices and compromises to tailor their instructional practices to their own personality and the local environment and constraints.  This goal has been aided by the participation of peer instructors and the practice of reflective journaling.  In an intermediate role between faculty and students, peer instructors can serve as 'spy' and sounding board, and alert faculty of student concerns outside of class time.  The journals allowed peer instructors and faculty to note areas of concern and conflict and reflect upon their causes and potential solutions.  There were numerous informal discussions among team members along these lines.  

We do not have definitive data to prove that instructors have completely developed this sort of knowledge, but can provide examples of strategies that were adopted in order to address student concerns.  In many cases these strategies did not involve changing practices dramatically, but simply reframing things with explicit attention to student expectations~\cite{Elby2010}.


Some instructors and curriculum developers might have concerns about the fidelity of adoption as instructors attempt to balance student expectations against instructional goals.  There is evidence to suggest that many faculty use curricula in ways the developers did not intend ~\cite{henderson2012, wieman2013, turpen2009}.  However, there is also evidence that as many as a third of faculty who adopt research-based instructional strategies discontinue their use often due to student resistance~\cite{henderson2012}.  It is our belief that instructors need the flexibility to make choices and the opportunity to reflect upon their practice with support from others.  

We quote Henderson \etal (2012):  
\begin{quotation}
Another way to think about this problem of discontinuance is in terms of support provided during implementation. While it may be possible to foresee implementation difficulties and provide faculty with additional advice before they begin to use a RBIS (Research-Based Instruction Strategy), it is almost certain that additional support during initial use will lead to more successful use. One important finding from a literature review on change strategies in higher education is that successful strategies provide support during implementation in the form of performance evaluation and feedback [87]. In current dissemination strategies, support and feed-back during implementation are quite rare. 
\end{quotation}

In our model of course reform implementation, there are three key factors that
may have supported the instructors' development of PCK and their recognition of mismatches in epistemological expectations. First, the
instructors had hands-on experience apply the reforms and refining their
implementation with the help of reflective journals that motivated the
instructors to think deeply about their teaching. The shared nature of the
journals helped open dialog amongst the faculty, peer instructors and
researchers. Second, faculty found that having a peer instructor present during
lecture meant having an person in the class who can both observe how the
students are engaging and how the faculty is teaching and interacting with the
class. This provided more feedback for the faculty in a timely fashion that was
critical in the faculty's ability to adapt the reform implementation for the
specific class. Third, the team of faculty, peer instructors and
discipline-based education researchers formed an open and supportive community
where expertise were shared and valued. Through this collaboration, the
researchers were able to bring in the appropriate research to help the faculty
implement the reforms more faithfully, the faculty were comfortable sharing
their reflections of their successes and frustrations which facilitated the team
in helping where they can, and the peer instructors were equal partners with
valuable insights and perspectives as student instructors.

\subsection{Content PCK} 

An important part of this process has been documenting our own students'
interactions with the course materials.  As the instructors interacted with
students in the classroom, they observed a number of recurring areas in which
students struggled, even with the strong support of the Lecture-Tutorials.  The
Peer Instructors recorded these areas of difficulty in their course notes and
journals and discussed them with the instructor after class; typically the Peer
Instructors would describe students struggling with an issue, sometimes
providing student quotes or estimates of clicker response rates.  Areas of
difficulty were discussed by the team informally, and the group worked to
document specific strategies that seemed to support student learning.  In this
way we have essentially formalized the process of PCK development. As part of
their assignment, the peer instructors categorized the different areas of
difficulty that most affected our classes, as shown in
Table~\ref{tab:difficulties}. In the section below, we provide additional detail
on selected examples as a means of illustrating the process.

We note that many of these examples of areas of conceptual difficulty have been
previously identified in the literature. However, our instructors, new to AER,
did not yet have sufficient background knowledge to easily identify and fully
make use of the AER community's findings during the semester. Thus we do not list
new findings, but rather issues that were particularly important in our
classrooms.   Our purpose in documenting these findings is to illustrate that the instructional team was able, through careful reflection, use of research-based curricular materials, and an ongoing commitment to listening to student reasoning, to develop a substantial amount of understanding of how students thought about course content and discern what aspects of the material were particularly challenging for students.  Instructors developed PCK in response to the specific challenges that
most affected our students and related to our desired curricular outcomes.  

\begin{table*}
\caption{\label{tab:difficulties} Example student difficulty areas related to introductory astronomy concepts that were identified during this work.}
\small
\begin{tabular}{p{4cm} p{5cm} p{8cm}}\hline
Topic & Context & Student difficulty with\ldots \\
\hline
Conceptual Reasoning & Newton's 3rd law and gravity & Accepting that two objects exert equal forces on each other, and that for three objects, there are two different equal and opposite forces acting. (LT, p.30)\\
& Absorption and emission & Identifying when a cloud of gas will be a source of light and when it will absorb light, particularly the sun's atmosphere, which is not "cold." (LT, p.63)\\
& Moon phases & Drawing the model, with half-lit up moons, and determining the fraction of the lit up side that would be visible from earth.\\
\hline
Quantitative Reasoning  & Kepler's 2nd law & Accepting that differently shaped triangles can have the same area. (LT, p.22)\\
& Parallax & Proportional reasoning and manipulations of the equation d = 1/p, and parallax angle as half angular separation. (LT, p.42)\\
\hline
Spatial Reasoning & Directions in celestial sphere & Identifying cardinal directions in celestial sphere and horizon views. (LT,p.3)\\
&Moon phases & Placing themselves within the model and identifying phases based on the position of the sun, moon, and earth (LT, p.81)\\
\hline
\end{tabular}
\end{table*}

\subsubsection{Examples of student difficulties}

\paragraph{Moon phases:} Students have difficulty visualizing the moon from the perspective of a person on the Earth.  

Studying the phases of the moon is known to be challenging for students, 
for example as reviewed in Trundle \emph{et al.}~\cite{Trundle2002} We found a number of issues specific
to the instructional materials and strategies used in our courses.  

\begin{figure*}
\includegraphics[width=0.7\linewidth]{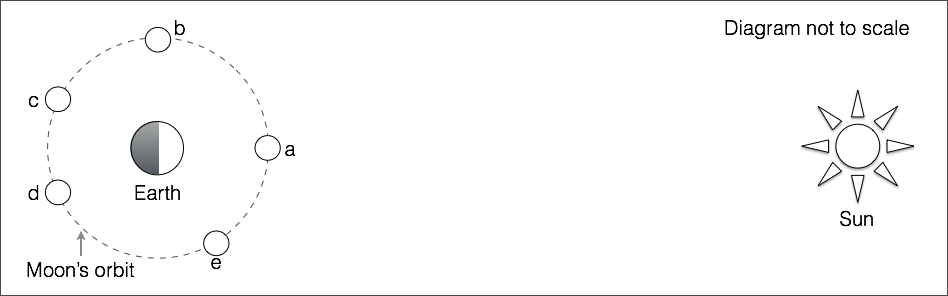}
\caption{Diagram of the Earth-sun-moon system that is used in class and on exams for questions concerning moon phases. Students might be asked to name the moon phase that occurs at one of the lettered moon positions, or to match a picture of a moon phase to the location in which it would occur.}\label{fig:moon}
\end{figure*}

In the Lecture-Tutorials and the lecture slides we introduce diagrams like the
one in Figure~\ref{fig:moon} to illustrate how one can determine
the phase of the moon from the position of the sun, moon, and earth. 
Both instructors and peer assistants observed that students attempt to memorize
the positions of different moon phases in the diagram; many students struggle when we pose questions
in which the arrangement of celestial bodies
differs from the initial presentation (for example, with the sun on the left instead of
the right as in Figure~\ref{fig:moon}.) 
To address this, we give students a succession of think-pair share questions
with similar images, but with the sun shining on the earth from different
directions and with the numbers and locations of the moon
positions varied.

During instruction, we also found that many students have a hard time imagining
how they would see the moon from the Earth perspective of this diagram, in order
to determine what fraction of the moon's visible surface would be illuminated by
the sun. We suggest that students use a method similar to that used in the lunar
phases module of the University of Nebraska's NAAP labs~\cite{NAAPlabs}: 
drawing a line from the earth to the center of the moon, and another line,
perpendicular to the first, bisecting the moon. The students are then asked
''what half of the moon can be seen from earth?'' and 
''how much and which side of it
is lit?''  As in many curricula, we have also found that demonstrations using a
light source as the sun and a ball as the moon help students to visualize moon
phases from different perspectives.  

\paragraph{Kepler's second law and triangle areas:} Many students believe that triangles of different shapes must have different areas.  

\begin{figure}
\includegraphics[width=0.5\linewidth]{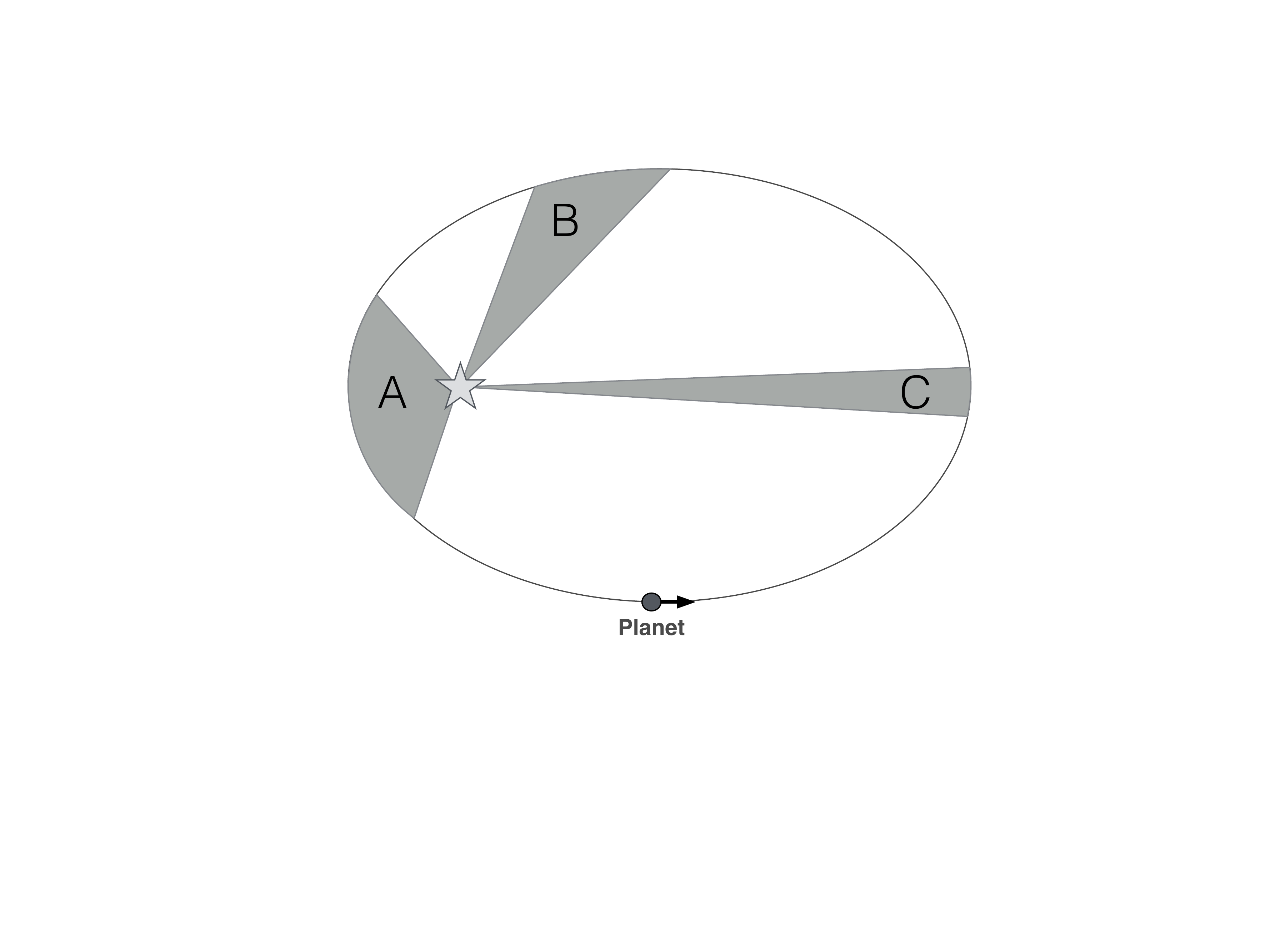}
\caption{Diagram of a planet in a Keplerian orbit around a star. The regions A, B, and C have the same area.}\label{fig:kepler}
\end{figure}

Two lecture periods of our course are devoted to Kepler's laws of planetary
motion.  
 In the Lecture-Tutorial and lecture slides related to Kepler's
second law, students are asked to refer to diagrams such as the one in
Figure~\ref{fig:kepler} and answer questions about the planet's motion, based on
the area of the regions that are swept out by the planet's orbit. For example,
students might be asked to compare the speed of the planet in regions B and
regions A, explicitly assuming that the areas of the regions are equal. 

We find that students have a hard time accepting that triangles with different
shapes can have equal areas, particularly those created by different
parts of an elliptical orbit, i.e., that a long, skinny triangle could have the
same area as a short, fat one. In addition, many students are reluctant to
attempt to estimate area, and feel frustrated about not being
presented with or being able to calculate an exact answer. Problems with
estimation and its role in science come up in other topics as well.  
We have additionally found that some students confuse the concepts of length and
area.  In orbital diagrams, students sometimes consider the length of the arc of
the orbit as opposed to the area swept out by the planet.

Sometimes using estimated triangle dimensions to calculate the areas and show
that they are close is useful.  One method we have found particularly
helpful in addressing these difficulties is to cut from the diagram two
differently shaped triangles that have the same area, and then tear one up into
pieces and show how the rearranged pieces fit into the other triangle.


\paragraph{Parallax:} There's math involved, therefore it must be hard. 

Throughout the course, several mathematical equations are used to illustrate
astronomical relationships. In the section of the course covering parallax, the
students are taught that the distance to an object in parsecs, $d$, is inversely
proportional to the parallax angle it makes as the Earth orbits the Sun, $p$,
presented with the equation, \begin{equation}
d = \frac{1}{p}.
\end{equation}
Many students immediately react negatively to this symbolic relationship,
perhaps because it is stated with unfamiliar words,
such as parallax, parsecs, and inversely proportional. We have also observed
that math in any form appears to intimidate and frustrate many students. It is
difficult to address this; we have not yet figured out a winning strategy. In
the midterm and final exams, questions related to parallax, are consistently
among the most missed. 

In addition to their objections to the formal mathematical form of this
relationship, some students struggle with the qualitative relationship
(inversely proportional to vs directly proportional).  They have a hard time
accepting that a star with a smaller distance has a larger parallax angle. To
help address this issue, we use participatory examples: for example, we ask
students to hold their thumb close to their faces, in front of their eyes, and
to close one eye, then the other, then extend their arm and repeat to show that
the angle through which their thumb jumps back and forth is larger when it was
closer. 


\paragraph{Celestial sphere and horizon diagrams:} Students struggle to visualize three-dimensional space when it is presented in a two-dimensional diagram.  

\begin{figure}
\includegraphics[width=0.4\linewidth]{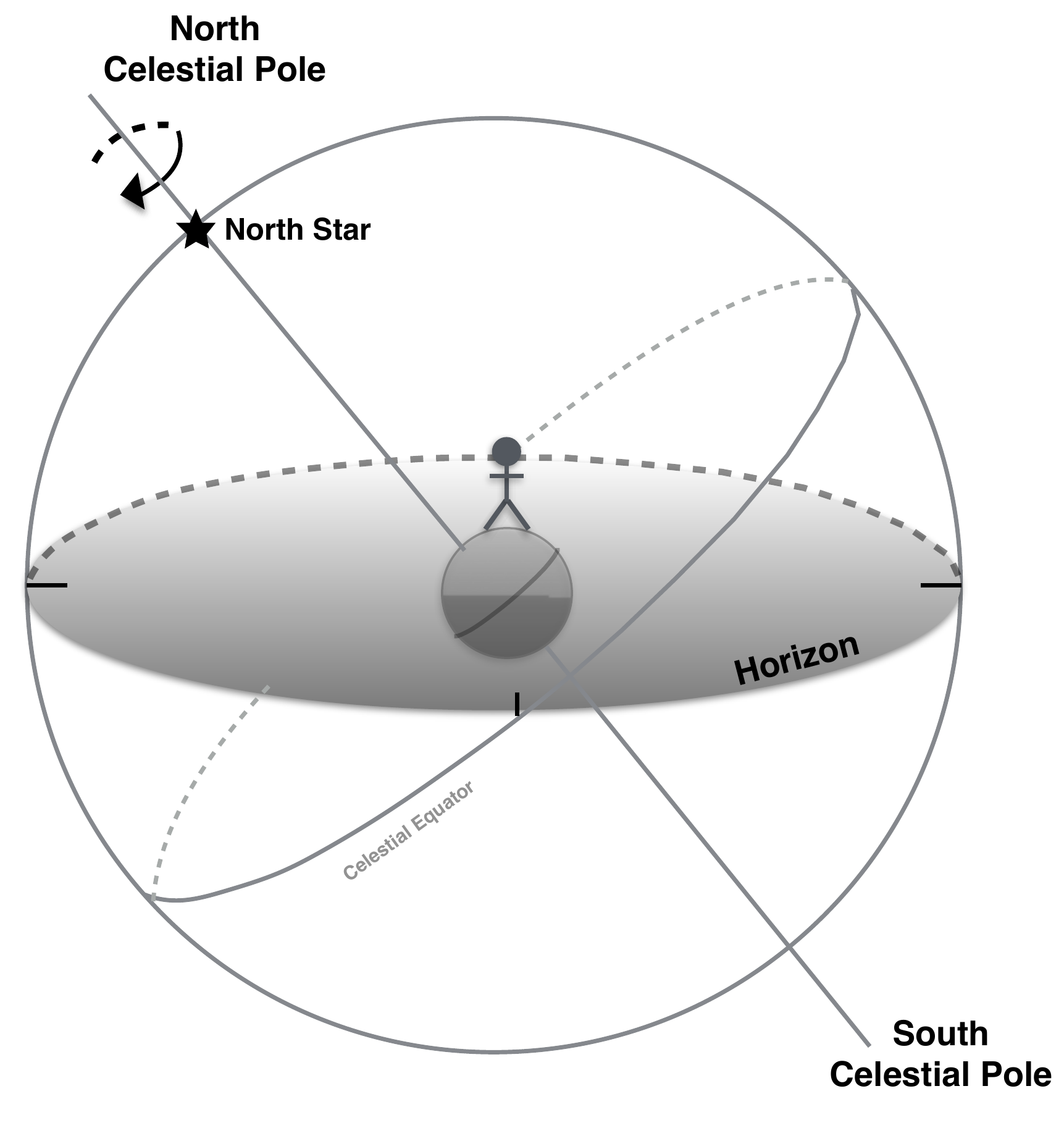}
\caption{Diagram of the celestial sphere showing an observer located at the mid lattitude of the northern hemisphere, and the location of the North star and the horizon. Students are asked to indicate the locations of the cardinal directions.}\label{fig:cs}
\includegraphics[width=0.4\linewidth]{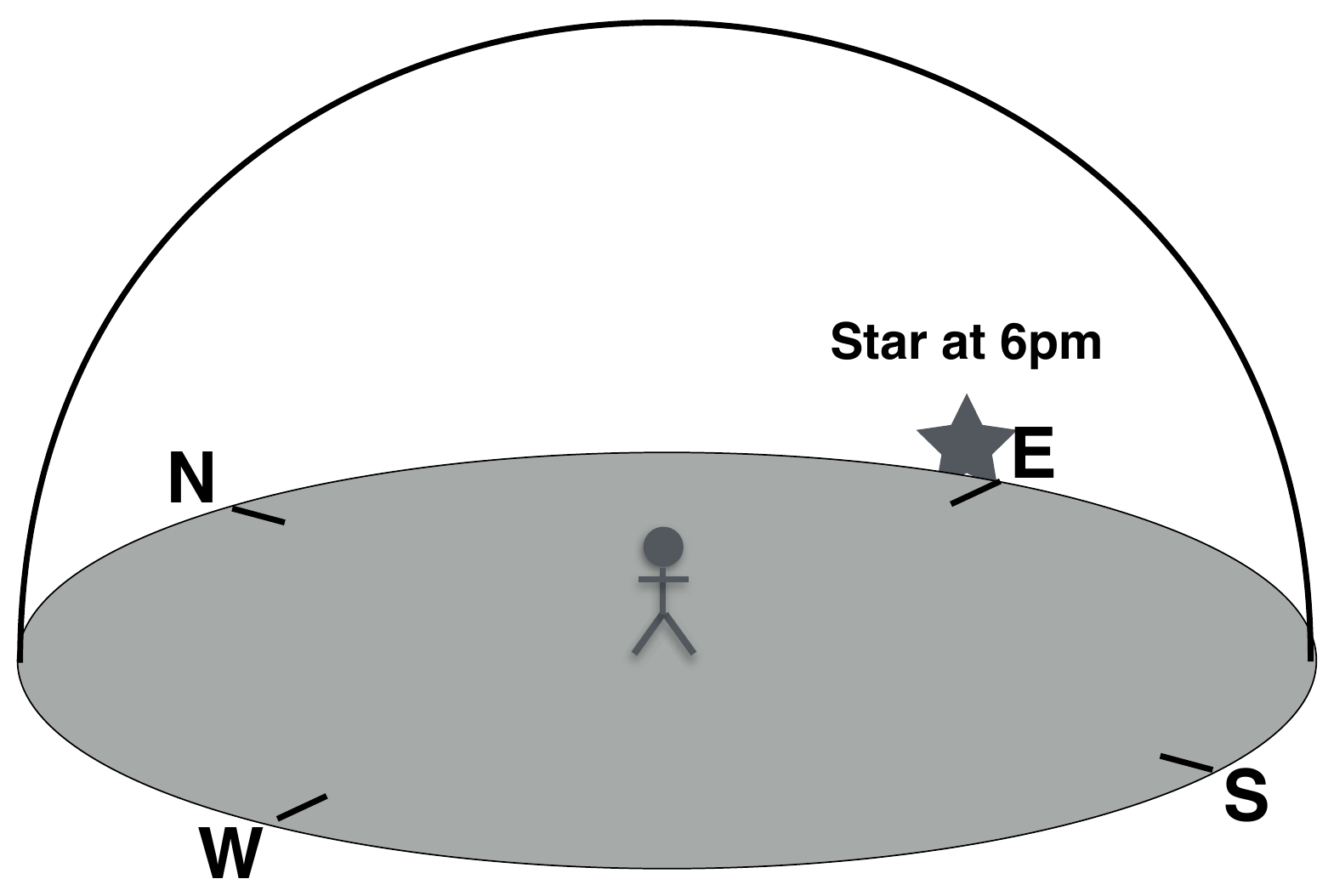}
\caption{Diagram of the celestial sphere showing the local sky with directions indicated. Students are asked to show the path that the star would take on the celestial sphere from where it rises to where it will set.}\label{fig:cs-perspective}
\end{figure}

The first two weeks of our courses heavily involve the use of two-dimensional
diagrams of the celestial sphere, some from the observer's perspective and
others from a perspective outside of the celestial sphere. We observed that
students have difficulty visualizing the three-dimensional space represented by
the two-dimensional printed figures. For example, when students are asked to
label the cardinal directions on a diagram like the one in Figure~\ref{fig:cs},
they label the cardinal direction North as near the North Star, South near the
``celestial sphere rotation" label, and East and West near the left and right of
the figure, near the horizon. Students also have difficulty determining the path
of a star around the dome as in Figure~\ref{fig:cs-perspective}. They often are
unable to visualize that the star would wrap over the surface of the dome to
connect east and west in the celestial sphere model. 

We tie the diagram to the three-dimensional
classroom setting by asking the students to point to cardinal directions
including North, and then to point in the direction of the North star.  We
emphasize the the difference between the two so students recognize that the
North star is up in the sky, but North is on the horizon---we sometimes remind
students that you couldn't drive North if it were up in the sky. We then relate
this exercise to the diagram:  for example,  North should be on the horizon
below the North Star, near star B's position 4.  To further exercise spatial
reasoning related to the celestial sphere model, we show animations that change
perspective from within the sphere to outside of it, and use physical models.
Physical celestial sphere models are often quite complicated and can add
additional confusion, but even simple transparent spheres marked with lines 
help with three-dimensional visualization. In the future we
will also incorporate the use of a planetarium session into our courses.

%% file: assessment.tex
The findings presented here span seven sections of Introduction to Astronomy at CSUF taught, as described in Section~\ref{sec:program}, by two junior faculty members, Smith and Read, from Fall 2011 until Spring 2013. One of the methods we used to assess student learning of the introductory astronomy concepts was analysis of pre- and post-tests administered as a multiple choice exam during the first and last weeks of class, respectively. We used a modified version of the Light and Spectroscopy Concept Inventory (LSCI) with the original 26 conceptual questions and a subset of the demographic questions~\cite{bardar2007}, which has a high conceptual overlap with the roughly one-fifth of our course that is focused on light.  

We quantified the student learning gain outcomes using the normalized gain~\cite{hake1998},
\begin{equation}
g = \frac{\mathrm{postscore}\%-\mathrm{prescore}\%}{100-\mathrm{prescore}\%}
\end{equation}
with $g$ being assessed on a student-by-student basis, using paired scores for only students that took both exams. 

From the outset, we had as a goal to implement interactive teaching strategies in our courses and to check our implementation, at least in the areas related to the LSCI, by comparing our normalized gains with those from national studies~\cite{rudolph2010}. Table~\ref{tab:lsci} shows the normalized gains measured from each section and the IAS determined by Rudolph based on data collected by the peer assistants and the instructors. These data all agree with the model-predicted gain from Rudolph \etal for ``all majors/areas of interest" as a function of IAS. Although the sampling is small, these data indicate that our local implementation of interactive learning methods and our development of PCK have resulted in courses with similar learning outcomes as those at other universities, even improving on nationally accepted standards.

\begin{table*}
\caption{\label{tab:lsci} Assessment scores for the course sections involved in this study. The \textit{prescore} and \textit{postscore} percentages are calculated from the scores from all $N$ students that took each exam. The normalized gain $g$ values were calculated using only paired scores from students who took both the pre- and post-exams. \textit{IAS} is the interactivity assessment score.}
\begin{tabular}{llllllll}\hline
Semester & Professor & Peer assistant & IAS \% & N & prescore \% & postscore \% & $g$ \%  \\
\hline
Fa2011 & Smith & Serna &  26.9 & 48 & 23.7 & 44.2 & 25.6 \\
Fa2012 & Smith & Serna & 39.4 & 38 & 22.2 & 50.5 & 36.7 \\
Fa2012 & Smith & - & 39.4 & 26 & 22.8 & 46.3 & 30.5 \\
Sp2013 & Smith & Chilton &41.9 & 31 & 26.3 & 55.2 & 40.4 \\
Sp2013 & Smith &  Chilton/Serna & 41.9 & 52 & 24.3 & 48.0 & 31.0 \\
Fa2012 & Read & Chilton & 34.7 & 33 & 22.0 & 47.2 & 31.8 \\
Sp2013 & Read & - &  37.7 & 40 & 23.2 & 48.6 & 32.4\\
\hline
\end{tabular}
\end{table*}

%% file: conclusions.tex
Ultimately we believe that we have achieved a successful and sustainable implementation of research-based instructional materials in the introductory astronomy course. The quantitative results described above are promising, and the qualitative results have helped us to further improve the course. The class has continued to grow and the instructors have not fallen off the implementation bandwagon; indeed, a more traditional instructor in the department has started to implement some of the strategies described.  In this section we reflect on what has made this successful implementation possible and what aspects of the project are generalization.

\subsection{Factors in success and generalizability}

Our successful implementation surely rests on a number of factors.  As we have reflected on this process, we have identified what we believe were key elements of our program that supported successful implementation.  As noted by prior research~\cite{henderson2009}, implementation often fails when faculty do not have sufficient support mechanisms and opportunities to reflect upon successes and difficulties.  

One factor that we believe has been important is the participation of a team including instructional faculty, peer instructors, and discipline-based education researchers.  Each team member brought perspectives and expertise to the table, and the team featured a very collegial atmosphere in which the contributions of all members were valued.  The instructional faculty who were implementing the modifications were the key drivers, but they welcomed the advice and insights of other team members.  The team also had access to a more extended network of participants in the Center for Astronomy Education workshop, particularly faculty members Alex Rudolph of Cal Poly Pomona and Duncan Brown of Syracuse University.  

We believe that the process of journaling was an important element of our program.  Each member of the team recorded journal entries and these proved to be an important mechanism for reflection as well as a repository of elements of PCK.  If we consider the categories of reform from Henderson and Dancy, this process involved the development of reflective instructors as well as the adoption and implementation of proven curriculum.  The process of reflection and discussion has proven particularly important for the adaptation of materials to the local setting and the individual instructors' personalities.  While the CAE materials and supporting instructional strategies are sound, the course instructors needed to make some choices that involved modifications to the standard practices proposed by CAE.  For example, as noted above, the instructors responded to student concerns about not receiving answers to tutorials by making more explicit the process of post-tutorial check-in.  

The local documentation of common areas of student difficulty was an important part of the process.  While the instructors had access to the instructor's guide and had participated in the CAE workshops, there were nevertheless many specific content areas where students struggled, and the exact nature of their difficulty was a surprise to instructors and not always documented in course materials.  The documentation of these areas of difficulty, including quotes from students, response patterns on clicker questions, and instructor questions that proved helpful, allowed the team to reflect on these areas of difficulty and to develop and test instructional strategies.  When one instructor found a particularly useful strategy or question, it could be shared with other members of the team.  

The peer instructors proved to be an exceptionally valuable part of the collaboration.  Not only did they assist in class with the implementation of course materials, but they also played a key role in documenting areas of student difficulty, both in terms of content and in terms of student expectations.  The peer instructors were exceptionally dedicated; the written journals from one student report ended up as a 90-page document including detailed class notes, clicker question results, and observations.

\subsection{Ongoing/future goals}

At present the modifications to the introductory course appear to be stable and sustainable.  Our goals for the future are to expand our efforts in several directions. Some can be characterized as expanding the modifications into new areas, and others as refining and sustaining the program for the introductory course.

The future goals for expansion take several forms.  The collaboration has already implemented a new laboratory course for the introductory course.  This new course was taught for the first time in Fall semester 2013.  Future goals include the introduction of intermediate-level courses suitable for physics majors and astronomy minors; the intention is to build these courses from the ground up with active learning strategies in mind.  Finally, several members of the team have expertise in gravitational-wave astronomy, and are testing instructional materials for this topic that could be suitable for either an intermediate-level course or for the introductory course.

We also hope to refine our efforts in the introductory course.  The process of collating and assembling the journals has allowed the team to reflect on a number of topics that students find challenging, and to share instructional strategies.  We have already made the compendium of these issues available to members of the collaboration, and feel that it might be of use to other instructors nationwide.  Finally, as our peer instructors have completed their bachelor's degrees and moved on to graduate study, we have a need for an infusion of new peer instructors.

%% file: ack.tex
This work was supported by NSF CAREER award PHY-1255650 and DOE FIPSE awards P116Z100226 and P116Z090274. The authors gratefully acknowledge the support of the College of Natural Sciences of Mathematics and the Department of Physics at California State University Fullerton, in particular Chair Jim Feagin and staff physicist Shovit Bhari. We thank Alex Rudolph for many fruitful discussions on this project and for helping us to calculate our interactivity scores. We thank Duncan Brown for course materials including lecture slides and exams and the NASA Center for Astronomy Education for their wonderful workshops, materials, and support.  